\documentclass[aps,prl,showpacs,twocolumn,superscriptaddress]{revtex4-2}%
\usepackage{times,mathptmx}

\usepackage{float}  %%%%%%%%%%%%%%%%%%%%%%%%%%%%%%%%%%%%%%%%%%%%%%%%%%%%%%%

\usepackage{graphicx}
\usepackage{graphics}
\usepackage[caption=false]{subfig}
\usepackage{color}
\usepackage[english]{babel}
\usepackage[utf8]{inputenc}
\usepackage{ae}
\usepackage{latexsym}
\usepackage{dcolumn}
\usepackage{bm}
\usepackage[table]{xcolor}
\usepackage{array}
\usepackage{url}
\usepackage{epsf}
\usepackage{epsfig}
\usepackage{pstricks,pst-grad}
\usepackage[pdftex,colorlinks]{hyperref}
\usepackage{amsmath}
\usepackage{amssymb}
\usepackage{ragged2e}
\usepackage{comment}

\begin{document}

\title{Understanding the Salt Effects on the Liquid-Liquid Phase Separation of Proteins}

\author{Chao Duan}
\affiliation{Department of Chemical and Biomolecular Engineering, University of California Berkeley, CA 94720, USA}

\author{Rui Wang}
\email {ruiwang325@berkeley.edu}\affiliation{Department of Chemical and Biomolecular Engineering, University of California Berkeley, CA 94720, USA}
\affiliation{Materials Sciences Division, Lawrence Berkeley National Lab, Berkeley, CA 94720, USA}

\date{\today}% It is always \today, today,

\begin{abstract}
Protein aggregation via liquid-liquid phase separation (LLPS) is ubiquitous in nature and intimately connects to many human diseases. Although it is widely known that the addition of salt has crucial impacts on the LLPS of protein, full understanding of the salt effect remains an outstanding challenge. Here, we develop a molecular theory which systematically incorporates the self-consistent field theory for charged macromolecules into the solution thermodynamics. The electrostatic interaction, hydrophobicity, ion solvation and translational entropy are included in a unified framework. Our theory fully captures the long-standing puzzles of the non-monotonic salt concentration dependence and the specific ion effect. We find that proteins show salting-out at low salt concentrations due to ionic screening. The solubility follows the inverse Hofmeister series. In the high salt concentration regime, protein remains salting-out for small ions but turns to salting-in for larger ions, accompanied by the reversal of the Hofmeister series. We reveal that the solubility at high salt concentrations is determined by the competition between the solvation energy and translational entropy of ion. Furthermore, we derive an analytical criterion for determining the boundary between the salting-in and salting-out regimes. The theoretical prediction is in quantitative agreement with experimental results for various proteins and salt ions without any fitting parameters. 
\end{abstract}

\maketitle

\section{Introduction}
% If your first paragraph (i.e. with the \dropcap) contains a list environment (quote, quotation, theorem, definition, enumerate, itemize...), the line after the list may have some extra indentation. If this is the case, add \parshape=0 to the end of the list environment.
Protein aggregation is ubiquitous in living cells, through which plenty of biomolecular condensates can be assembled \cite{Alberti2021,Lyon2021}. These biomolecular condensates play a vital role in cellular organization and functions, such as the formation of nucleoli \cite{Lafontaine2021}, heterochromatin and ribonucleoprotein granule \cite{Brangwynne2009,Strom2017} as well as signal transduction within the cytoplasm \cite{Pawson2010,Chong2016,Woerner2016}. In addition, the aggregation of various misfolded proteins intimately linked to many neurodegenerative diseases including Alzheimer’s, Parkinson’s, diabetes, and prion diseases \cite{Knowles2014,Chiti2017}.
Evidence is mounting that protein aggregation proceeds via a liquid-liquid phase separation (LLPS), which is manifested as the formation of a dense phase often resembling liquid droplets and a coexisting dilute phase \cite{Ray2020,Riback2020,Fuxreiter2021,Shimobayashi2021}.
Revealing the essential physical chemistry of the LLPS-driven aggregation will help delineate the functions of biomolecular condensates and provides useful guidance for the therapy of diseases \cite{Brangwynne2015,Shin2017,Choi2020}. In spite of increasing academic interests, understanding and regulating LLPS of protein remains a big challenge \cite{Alberti2019}.

Salt effect on LLPS of protein is one of the most long-standing puzzles. It is well-known that the ionic environment has critical impacts on the LLPS; besides, the addition of salt also provides an effective tool to modulate it \cite{Posey2018}. However, this salt effect is very complicated: the LLPS of protein has non-trivial dependence on both the salt concentration and the chemical identity of ions (usually known as the specific ion effect) \cite{Zhang2010,Okur2017,Wei2017,Krainer2021,Rogers2022}. Zhang and Cremer measured the cloud point of positively-charged lysozyme solutions \cite{Zhang2009}. At low salt concentrations, they found that the solubility of lysozyme decreases as salt concentration increases, i.e. protein salting-out. The increase of solubility follows the inverse Hofmeister series of anion. In contrast, at high salt concentrations, lysozyme remains to show salting-out for some anions (e.g. Cl$^{-}$), whereas other anions (e.g. Br$^{-}$ and I$^{-}$) enhance the lysozyme solubility, i.e. protein salting-in. The solubility increase follows the direct Hofmeister series in the high salt concentration regime. Neither the non-monotonic salt concentration effect nor the specific ion effect can be explained, even qualitatively, by the standard mean-field Poisson-Boltzmann (PB) theory \cite{Ben-Yaakov2009}. Similar salt-dependent behaviors have also been observed in other protein solutions \cite{Cho2008,Mason2010,Patel2017,Otis2022} and soft matter systems such as synthetic polymers \cite{Zhang2005,Weckstrom2007,Shimada2011} and colloidal dispersions \cite{Stradner2004,Salis2014}, implying the universality of the salt effects on LLPS.

\begin{figure*}[t]
\centering
\includegraphics[width=0.9\textwidth]{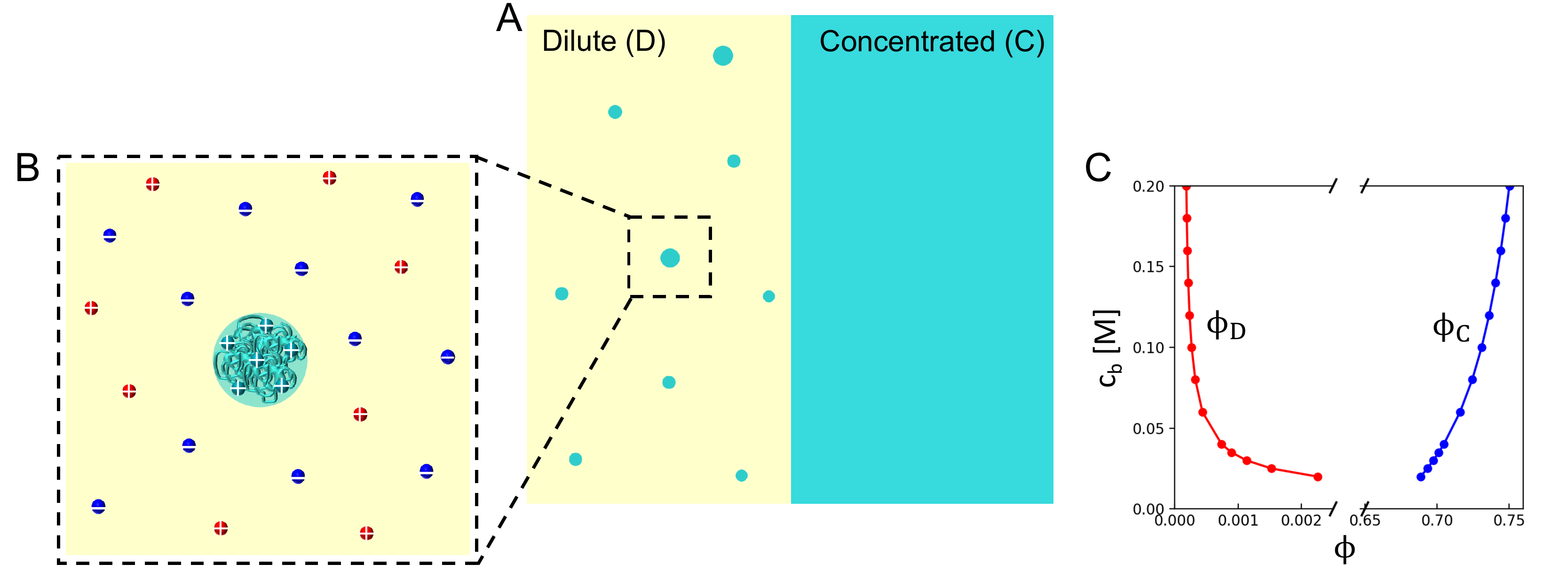}
\caption{({\it A}) Schematic of the total system consisting of coexisting dilute phase (D) and concentrated phase (C). The dilute phase is an assemble of protein aggregates with different aggregation number. ({\it B}) A subsystem containing one isolated aggregate in the presence of salt ions.
({\it C}) A representative phase diagram plotting the equilibrium volume fractions of the two coexisting phases ($\phi_D$ and $\phi_C$) as a function of bulk salt concentration $c_b$. $a_+=a_-=2.5\mathring{\rm A}$, $z_+=z_-=1$, $\epsilon_P=30$, and $\epsilon_S=80$.}
\label{fig1}
\end{figure*}

Many theoretical and computational efforts have been made to explain these salt effects. Kastelic et al. assumed a phenomenological model for the interaction energy between proteins, where the well depths in the presence of different alkali-halide salts were fitted to experimental data \cite{Kastelic2015}. They suggested that the salt effect on LLPS is majorly attributed to the ionic screening, but the salting-in behavior and the reversal of Hofmeister series observed at high salt concentrations have not been captured. Zhang and Cremer developed a modified binding isotherm model \cite{Zhang2009}. The model parameters representing the effectiveness and equilibrium constant for the association of a specific anion to protein surface were fitted to the measured cloud point. They found that the salt effect in the high salt concentration regime is correlated to the interfacial tension of protein surrounded by anions with different polarizability. Furthermore, using a modified PB theory to account for ion size and polarizability, Bostr$\ddot{\rm o}$m et al. suggested that the reversal of the Hofmeister series at high salt concentrations originates from the inversion of effective surface charge of proteins \cite{Bostrom2011}.
However, there is no theory up to now that can unify the description of the salt effects on LLPS of proteins for the entire salt concentration regime. The underlying physical chemistry, particularly for the counterintuitive behaviors observed at high salt concentrations, is still unclear.      

To uncover the salt effect on LLPS of protein, we develop a molecular theory which systematically includes the electrostatics, hydrophobic interaction, ion solvation and translational entropy of protein in a unified framework. Compared to the existing theories, we have made the following two major improvements. First, we explicitly account for the highly localized density fluctuation of proteins in the dilute phase rather than assuming random mixing as invoked in the Flory-Huggins (F-H) theory \cite{Wang2012,Duan2023}. This enables the accurate treatment of ionic screening effect on a charged protein aggregate. Second, we include the self-energy of ions as a result of electrostatic fluctuation, which captures the salt effects beyond the mean-field PB level \cite{Wang2010,Agrawal2022}. Our theory predicts that protein salting-out at low salt concentrations is attributed to the screening effect, whereas protein solubility at high salt concentrations is determined by the competition between the solvation energy and translational entropy of ions. Furthermore, we derive an analytical criterion for determining the boundary between the  salting-in and salting-out regimes for different proteins and ions. The theoretical prediction is in quantitative agreement with experimental data reported in literature without any fitting parameters.

\section{Theory}

The solubility of protein in a salt solution is built upon the equilibrium between a dilute phase and a protein-rich concentrated phase as illustrated in Fig. 1{\it A}. The concentrated solution can be modeled by a homogeneous liquid-like condensate due to the negligible density fluctuation and the surface contribution. However, the description of the dilute phase is nontrivial because of the large localized density fluctuation. An instantaneous picture of the dilute protein solution has localized high concentrations where the proteins are located and pure salt solutions elsewhere. This is an exactly different scenario compared to that envisioned in the random mixing picture of F-H theory used in existing work \cite{Brangwynne2015,Shin2017,Choi2020,Martin2020,Gouveia2022}. To account for this large localized density fluctuation in the dilute phase, we focus on the subvolume of the entire solution containing only one isolated protein or one multi-protein aggregate (see Fig. 1{\it B}). The density profile and free energy of the protein/aggregate is obtained by applying the self-consistent field theory (SCFT) in the subvolume. This information is then incorporated into the framework of dilute solution thermodynamics to reconstruct the solution behavior of the entire dilute phase.

\subsection*{Self-Consistent Field Theory for an Isolated Protein/Aggregate} As shown in Fig. 1{\it B}, we consider a subvolume consisting of an isolated aggregate of $m$ proteins and $n_S$ solvent molecules in the presence of $n_{\pm}$ mobile ions with valency $z_{\pm}$. $m=1$ specifies the case of an isolated protein. The subvolume is taken as a semicanonical
ensemble: the number of proteins is fixed whereas solvent
and mobile ions are connected with a bulk salt solution of
ion concentration $c^b_{\pm}$ that maintains the chemical potentials of the solvent $\mu_S$ and ions $\mu_{\pm}$ \cite{Duan2020,Duan2023}. The proteins considered here are assumed to be unfolded or intrinsic disordered, where the widely-adopted charged macromolecular model is invoked to describe these proteins \cite{Choi2020,Mezzenga2013,vanderLee2014}. This model is also general for synthetic polyelectrolytes and other biomacromolecules \cite{Muthu2017}. The charged macromolecule is assumed to be a Gaussian chain of $N$ Kuhn segments with Kuhn length $b$.
The smeared charge model is adopted to describe the backbone charge distribution with the charge density $\alpha$ \cite{Fredrickson2006}. For For simplicity, the volume of the chain segment and the solvent molecule
are assumed to be the same $v_0$. The local hydrophobic interaction between protein and solvent is described by the Flory parameter $\chi$.  The key results of the SCFT are the following set of equations for protein density $\rho_P({\bf r})$, conjugate
fields $\omega_P(\bf r)$ and $\omega_S(\bf r)$, electrostatic potential $\psi({\bf r})$ and
ion concentration $c_{\pm}({\bf r})$ (see SI, Section 1 for the detailed derivation):
\begin{align}
\omega_P({\bf r})&-\omega_S({\bf r}) = \chi[1-2\rho_P({\bf r})] - \frac{v_0}{2}\frac{\partial \epsilon({\bf r})}{\partial \rho_P({\bf r})}[\nabla\psi({\bf r})]^2  \nonumber\\
& +\alpha\psi({\bf r}) + v_0 \left [ c_+({\bf r}) \frac{\partial u_+({\bf r}) }{\partial \rho_P({\bf r})} + c_-({\bf r}) \frac{\partial u_-({\bf r}) }{\partial \rho_P({\bf r})} \right ]   \tag{1a}
\end{align}
\begin{align}
\rho_P({\bf r}) &= \frac{m}{Q_P}\int_0^N {\rm d}sq({\bf r},s)q({\bf r},N-s)   \tag{1b}
\end{align}
\begin{align}
1-\rho_P({\bf r}) &= {\rm e}^{\mu_S}{\rm exp}[-\omega_S({\bf r})]   \tag{1c}
\end{align}
\begin{align}
-\nabla\cdot[\epsilon({\bf r})\nabla\psi({\bf r})] &=z_+ c_+({\bf r})-z_- c_-({\bf r})+\frac{\alpha}{v_0}\rho_P({\bf r})  \tag{1d}
\end{align}
\begin{align}
c_{\pm}({\bf r}) &=\lambda_{\pm} {\rm exp}[\mp z_{\pm} \psi({\bf r})-u_{\pm}({\bf r})]   \tag{1e}
\end{align}
where $\epsilon({\bf r})=kT \epsilon_0 \epsilon_r({\bf r}) /e^2$ is the scaled permittivity with $\epsilon_0$ the vacuum permittivity,
$e$ the elementary charge and $\epsilon_r({\bf r})$ the local dielectric constant. $\epsilon_r({\bf r})$ can be evaluated based on the local composition \cite{Wang2011,Sing2012}. Here a linear mixing rule is adopted which leads to $\epsilon_r({\bf r})=\epsilon_P \rho_P({\bf r})+\epsilon_S (1-\rho_P({\bf r}))$, with $\epsilon_P$ and $\epsilon_S$ the dielectric constant of the pure protein and solvent, respectively \cite{Wang2011,Sing2012,Zhuang2021}. $\lambda_{\pm}=e^{\mu_{\pm}}/v_{\pm}$ is the fugacity of the ions controlled by the bulk salt concentration.
$Q_P$ is the single-chain partition function given by $Q_P=(1/v_0)\int {\rm d}{\bf r}$ $q({\bf r},N)$, whereas $q({\bf r},s)$ is the chain
propagator determined by the diffusion equation
\begin{align}
\frac{\partial q({\bf r},s)}{\partial s} &= \frac{b^2}{6}\nabla^2q({\bf r},s)-\omega_P({\bf r})q({\bf r},s)  \tag{2}
\end{align}

$u_{\pm}(\bf r)$ in Eq. 1e in the self energy of ions resulting from the fluctuation of the electrostatic field \cite{Wang2010,Agrawal2022}. If the nonuniversal
contribution of the fluctuation in the length scale of the ion size is retained, $u_{\pm}(\bf r)$ reduces to the local Born energy as:
\begin{align}
u_{\pm}({\bf r}) =& \frac{z_{\pm}^2e^2}{8\pi a_{\pm} \epsilon({\bf r})}  \tag{3}
\end{align}
with $a_{\pm}$ the Born radius of ions. The Born solvation energy accounts for the electrostatic interaction between the ion and the local dielectric medium \cite{Wang2011,Wang2014}. It captures the fact that ions are more preferable to be distributed in the medium with higher dielectric constant. For systems with spatially
varying dielectric permittivity, $u_{\pm}$ is not a constant, and cannot be adsorbed into the
redefinition of the chemical potential. It will thus affect both the ion distribution and protein density profile as indicated in Eqs. 1a and 1e. The non-local contributions of electrostatic fluctuation, such as ion correlation and image force, can be rigorously included into the self energy through Gaussian variational approach \cite{Wang2010,Agrawal2022}. We refer interested readers to the relevant literature for more details. The free energy of the subsystem is then
\begin{align}
F_m &= -m{\rm ln}Q_P+{\rm ln}(m!)-{\rm e}^{\mu_S}Q_S \nonumber\\
&+\frac{1}{v_0}\int {\rm d}{\bf r} \left [ \chi\rho_P (1-\rho_P)
-\omega_P\rho_P-\omega_S (1-\rho_P) \right] \nonumber\\
&+\int {\rm d}{\bf r} \left [ \frac{\alpha}{v_0}\rho_P\psi-\frac{\epsilon}{2}(\nabla\psi)^2-c_+ -c_- +c^b_+ +c^b_-\right ] \tag{4}
\end{align}

\subsection*{Phase Equilibrium}
The protein solution in the dilute phase can be reconstructed by incorporating the density profile and free energy of the $m$-aggregate obtained from SCFT into the framework of dilute solution thermodynamics \cite{Duan2023}. The free energy density of the entire dilute solution with volume $V$, including the translational entropy of aggregates, can be written as
\begin{align}
\frac{F_D}{V}=&
\sum_{m=1}^{\infty} \{C_m F_m + C_m [{\rm ln} (C_m v_m)-1] \}  \tag{5}
\end{align}
where $C_m$ is the concentration of the $m$-aggregate. $v_m$ is a reference volume which for simplicity can be taken as the volume of the $m$-aggregate. $C_m v_m$ thus becomes the corresponding volume fraction $\phi_m$ of the $m$-aggregate.  In Eq. 5, the interaction between different aggregates is ignored under the assumption of sufficiently dilute solution. The equilibrium concentration of $m$-aggregate can be obtained by minimization of the free energy density in Eq. 5 subject to fixed total protein concentration $\sum^{\infty}_{m=1} mC_m$, which results in the following distribution:
\begin{align}
\phi_m=& \phi^m_1 {\rm exp}(-\Delta F_m) \tag{6}
\end{align}
Here $\Delta F_m=F_m-mF_1$ is the free energy of formation of the $m$-aggregate from $m$ single isolated proteins.

The protein solution in the concentrated phase can be modeled as an infinitely large aggregate with uniform protein density. The free energy density is directly obtained by applying SCFT to a homogeneous system and Eq. 4 becomes: 
\begin{align}
\frac{F_C}{V} &= \frac{\phi_P}{N}{\rm ln}\left(\frac{\phi_P}{N}-1 \right) + (1-\phi_P)[{\rm ln}(1-\phi_P)-1)]  \nonumber\\  
&+ \chi\phi_P (1-\phi_P) + \frac{\alpha}{v_0}\phi_P \psi - c_+ - c_- + c_+^b + c_-^b  \tag{7}
\end{align}
where $\psi$ is the electrostatic potential difference between the concentrated phase and the dilute phase usually known as Donnan potential or Galvani potential \cite{Agrawal2022,Wang2011}. $\psi$ is obtained by applying the charge neutrality constrain to the homogeneous concentrated phase.

The equilibrium between the protein dilute phase and the protein concentrated phase is determined by the respective equality of the chemical potential of the protein and the solvent in the two coexisting phases, which results in
\begin{align}
F_1+{\rm ln}\phi_1 &={\rm ln}\frac{\phi_C}{N}-1+(1-N)(1-\phi_C)+\chi N(1-\phi_C)^2 \nonumber \\
&+{\mu}^{elec}_P  \tag{8a}
\end{align}
\begin{align}
-\sum_{m=1} \frac{\phi_m}{mN} &=\left(1-\frac{1}{N}\right)\phi_C+{\rm ln}(1-\phi_C)+\chi N{\phi_C}^2+{\mu}^{elec}_S  \tag{8b}
\end{align}
where $\phi_C$ is the equilibrium volume fraction of protein in the concentrated phase and $\phi_m$ is the equilibrium volume fraction of the $m$-aggregate in the dilute phase given by Eq. 6. The total volume fraction of protein in the dilute phase is thus $\phi_D = \sum_{m=1}\phi_m$.
It should be noted that the sum on the left-hand side of Eq. 8b is the dimensionless osmotic pressure in dilute phase (in accordanace with Van't Hoff law) as expected for an ideal solution \cite{Doi2013}.
${\mu}^{elec}_P$ and ${\mu}^{elec}_S$ are the electrostatic contributions in the chemical potentials of protein and solvent, respectively, which are given by
\begin{align}
{\mu}^{elec}_P &= \alpha N \psi - Nv_0[(c_+-c^b_+)+(c_--c^b_-)] \nonumber\\
&+Nv_0(u_+c_++u_-c_-)\left(\frac{\epsilon_S-\epsilon_P}{\epsilon}\right)(1-\phi_C)  \tag{9a}
\end{align}
\begin{align}
{\mu}^{elec}_S &=-v_0[(c_+-c^b_+)+(c_--c^b_-)] \nonumber\\
&+v_0(u_+c_++u_-c_-)\left(\frac{\epsilon_S-\epsilon_P}{\epsilon}\right)\phi_C  \tag{9b}
\end{align}
It is worth noting that the three terms on the right hand side of Eq. 9a represent the contributions from the energy of a charged protein in the electrostatic field, translational entropy and the solvation energy of salt ions, respectively. For each salt concentration $c_b$ in the bulk salt solution (i.e. reservoir), the equilibrium volume fractions in the coexisting dilute and concentrated phases $\phi_D$ and $\phi_C$ are obtained by  solving Eqs. 8a and 8b simultaneously, from which the phase diagram as illustrated in Fig. 1{\it C} can be obtained.

\section{Results}

In the current work, we focus on the salt concentration effect and the specific ion effect. The number of Kuhn segments in the protein is set as $N=50$ with $b=1.0$nm. We use the simple system of a homogeneous chain with uniform backbone charge distribution to illustrate the fundamental physical chemistry. The backbone charge density $\alpha=+0.05$, where positive $\alpha$ is adopted to facilitate the comparison with the corresponding proteins studied in experiments \cite{Zhang2009,Cho2008,Mason2010,Otis2022}. The volume of the chain segment and the solvent molecule is assumed to be the same as $v_0=1.0$nm$^3$. The temperature is set to be 298K with the Flory parameter $\chi=1.2$.
%The numerical details are provided in {\it SI Appendix}, Section 2.

\begin{figure}[h]
\includegraphics[width=0.45\textwidth]{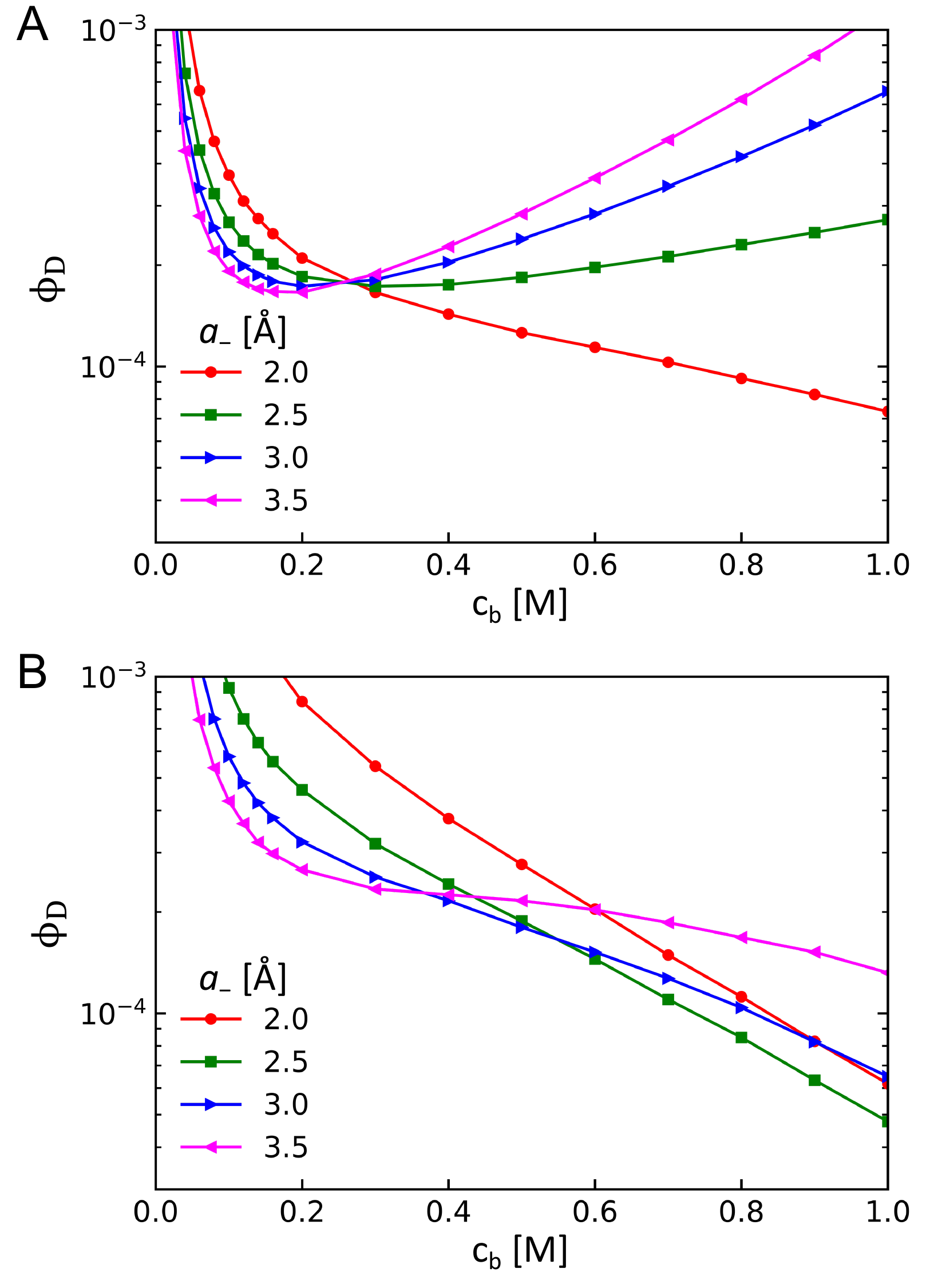}
\caption{Protein solubility $\phi_D$ as a function of salt concentration $c_b$ for anions with different radius $a_{-}$. $a_+=2.5\mathring{\rm A}$, $z_+=z_-=1$, $\epsilon_S=80$. ({\it A}) $\epsilon_P=30$ and ({\it B}) $\epsilon_P=10$.
}
\label{fig2}
\end{figure}

\subsection*{Salt Effects on the Protein Solubility} The salt effects on LLPS of protein observed in experiments show complicated dependence on both the salt concentration and the chemical identity of ions. We theoretically investigate the protein solubility  for different salt concentrations and various anion radii. Here, the solubility is represented by $\phi_D$, the equilibrium volume fraction of the dilute phase on the coexistence curve (see Fig. 1{\it C}). Figure 2{\it A} shows that the solubility decreases as $c_b$ increases in the low salt concentration regime ($c_b<0.2$M), indicating protein salting-out. At the same $c_b$, the solubility decreases with the increase of anion radius, consistent with the trend of inverse Hofmeister series. In contrast, in the high salt concentration regime ($c_b>0.2$M), protein remains salting-out for small anion ($a_-=2.0\mathring{\rm A}$), but turns to salting-in for larger ions. The solubility increases with the increase of anion radius, indicating the direct Hofmeister series. The dependence of LLPS on both the salt concentration and the specific ions predicted by our theory is in good agreement with the solubility measurements of lysozyme in Zhang and Cremer's experiments \cite{Zhang2009}. Particularly, they found salting-out behavior at high salt concentrations only for small Cl$^{-}$, whereas all other larger anions show salting-in behavior, exactly captured by Fig. 2{\it A}.

\begin{figure*}[t]
\centering
\includegraphics[width=0.9\textwidth]{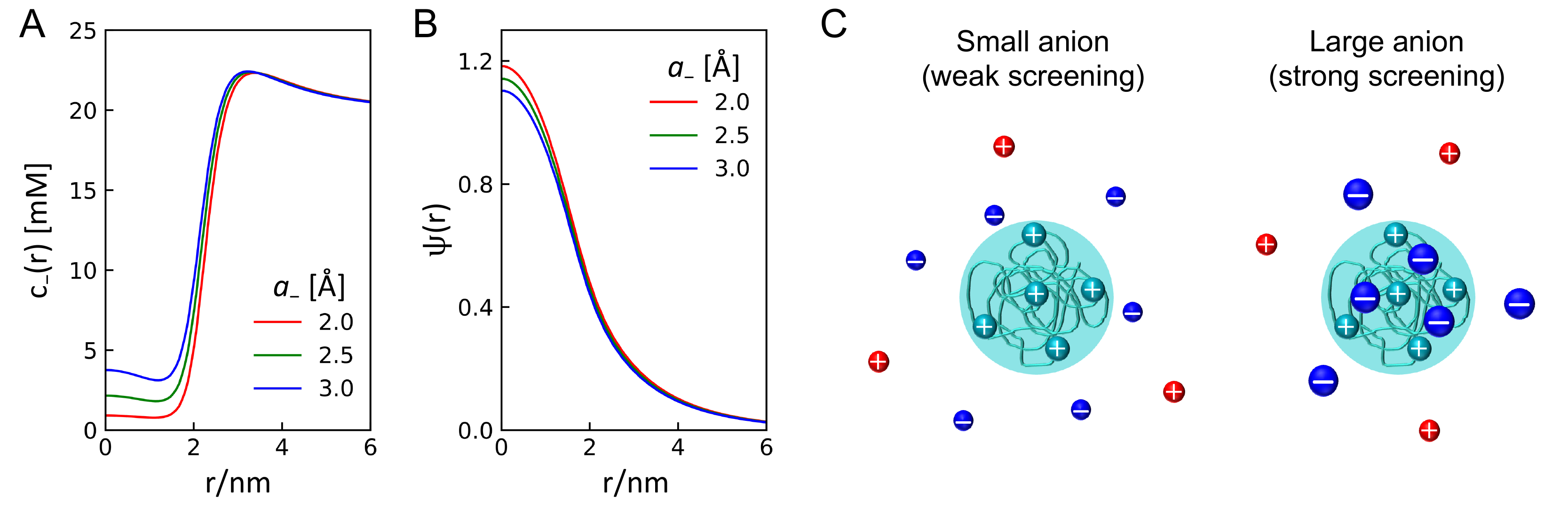}
\caption{The salt effect on the electrostatic double layer structure around a positively-charged protein in the low salt concentration regime. ({\it A}) anion concentration profile $c_-(r)$ and ({\it B}) electrostatic potential $\psi(r)$. $a_+=2.5\mathring{\rm A}$, $z_+=z_-=1$, $\epsilon_P=10$, $\epsilon_S=80$, and $c_b=20$mM. ({\it C}) Schematics of different screening effects for small anion and large anion.}
\label{fig3}
\end{figure*}

The salt effects on the solubility in the high salt concentration regime also depends on the property of protein. If a protein with lower dielectric constant ($\epsilon_P=10$) is adopted as shown in Fig. 2{\it B}, it exhibits salting-out in the entire salt concentration regime for all the anions with $a_- \le 3.5\mathring{\rm A}$.
This is in stark contrast with the behavior predicted for proteins with high $\epsilon_P$. It is interesting to note that the same trend has also been reported in experiments. Cho et al. measured the solubility of elastin-like polypeptide which has lower dielectric constant than lysozyme \cite{Cho2008}. All anions investigated in their work show salting-out at high salt concentrations. 
Similar all salting-out behavior has been observed by Zhang et al. in the synthetic poly(N-isopropylacrylamide)  (PNIPAM) system\cite{Zhang2005}. The dielectric constant of PNIPAM is less than 5 as reported in the literature \cite{Rullyani2020}.
These experimental results are in good agreement with our theoretical prediction.

As elucidated in Eq. 8, the LLPS of protein is determined by the interplay between the hydrophobic attraction, Coulomb repulsion, as well as the solvation energy and the translational entropy of ions. The solubility is directly controlled by the effective two-body interaction between proteins: attractive contributions to the interaction favor condensation, whereas repulsive contributions prefer dissolution. The impacts of the aforementioned four contributions on the two-body interaction and their salt-concentration dependence are summarized in Table 1. The hydrophobicity of protein backbone always leads to effective attraction and is independent of $c_b$, which thus can be neglected when considering salt effects. Coulomb interaction between likely-charged proteins is repulsive and decays exponentially with $c_b$ as a result of ionic screening. Furthermore, the contribution of ion solvation is effectively attractive. Ions prefer to be dissolved in the medium with higher dielectric constant as indicated by the Born solvation model (Eq. 3). This selective partition leads to depletion of ions from proteins and thus drives phase separation. Lastly, the translational entropy of ions favors a uniform distribution in the entire solution, which suppresses the aggregation of proteins and thus provides an effective repulsion. As illustrated in Eq. 9, the contributions of both the ionic solvation and translational entropy depend linearly on $c_b$. In the following two subsections, we will provide more detailed analysis on the salt effects in the low and high salt concentration regimes, respectively.

\begin{table}[h]%The best place to locate the table environment is directly after its first reference in text
\caption{\label{tab:table1}%
The impacts of different contributions and their salt-concentration dependence on the effective two-body interaction between proteins
}
\begin{ruledtabular}
\begin{tabular}{lcc}
%\textrm{Left\footnote{Note a.}}&
%\textrm{Centered\footnote{Note b.}}&
Contribution & Effective interaction & $\rm c_b$-dependence \\
\colrule
hydrophobicity & attractive & $\sim c^0_b$ \\
Coulomb interaction & repulsive & $\sim e^{-\kappa r}/r$ ($\kappa \sim c^{1/2}_b$)  \\
ion solvation & attractive & $\sim c_b$ \\
entropy of ion & repulsive & $\sim c_b$ \\\end{tabular}
\end{ruledtabular}
\end{table}

\subsection*{Ionic Screening at Low Salt Concentrations} In the low salt concentration regime, the Coulomb repulsion between proteins dominates compared to the contributions from ionic solvation and translational entropy. Thus, the key factor that determines the salt effects on LLPS is how the Coulomb repulsion are screened by salt ions. The screening effect gets stronger as $c_b$ increases, which leads to the reduction of the effective charge of protein and thus weakens the two-body repulsion. Therefore, the solubility of protein decreases as $c_b$ increases, indicating a salting-out behavior (see Fig. 2).

While the salting-out behavior is universal for all ions in the low salt concentration regime, its degree exhibits specific ion effect because of the different efficacy of anions in screening the Coulomb repulsion. Based on the Born solvation model, ions are more preferable to be distributed in the solvent region than the protein region as $\epsilon_S>\epsilon_P$ in most cases. This selective partition becomes more pronounced for smaller anions. Figure 3 shows the electrostatic double layer structure around a positively-charged protein. Anions with smaller radius are repelled more from the protein center, resulting in a less screened Coulomb potential. Therefore, protein solubility decreases with the increase of the anion radius, in agreement with the trend of inverse Hofmeister series observed in experiments at low salt concentrations. Zhang and Cremer suggested that the specific ion effect on LLPS in the low salt concentration regime is mainly originating from the effectiveness of anions with different sizes in associating with the positively-charged protein \cite{Zhang2009}. Their explanation is consistent with the mechanism revealed in our results.

\subsection*{Competition between Ion Solvation and Translational Entropy at High Salt Concentrations} In the high salt concentration regime, the charges carried by proteins are largely screened, and hence the Coulomb repulsion becomes less significant. The LLPS of protein is mainly determined by the competition between the solvation and translational entropy of ions as illustrated in Fig. 4. The tendency for ions to be preferentially solvated by medium with higher dielectric constant leads to a driving force for the separation of proteins from the solvent phase. This reduces the solubility, i.e. salting out. On the contrary, the translational entropy of ions favors a uniform distribution in the entire system, which enhances the miscibility between protein and solvent, i.e. salting in.

\begin{figure}[h]
\includegraphics[width=0.49\textwidth]{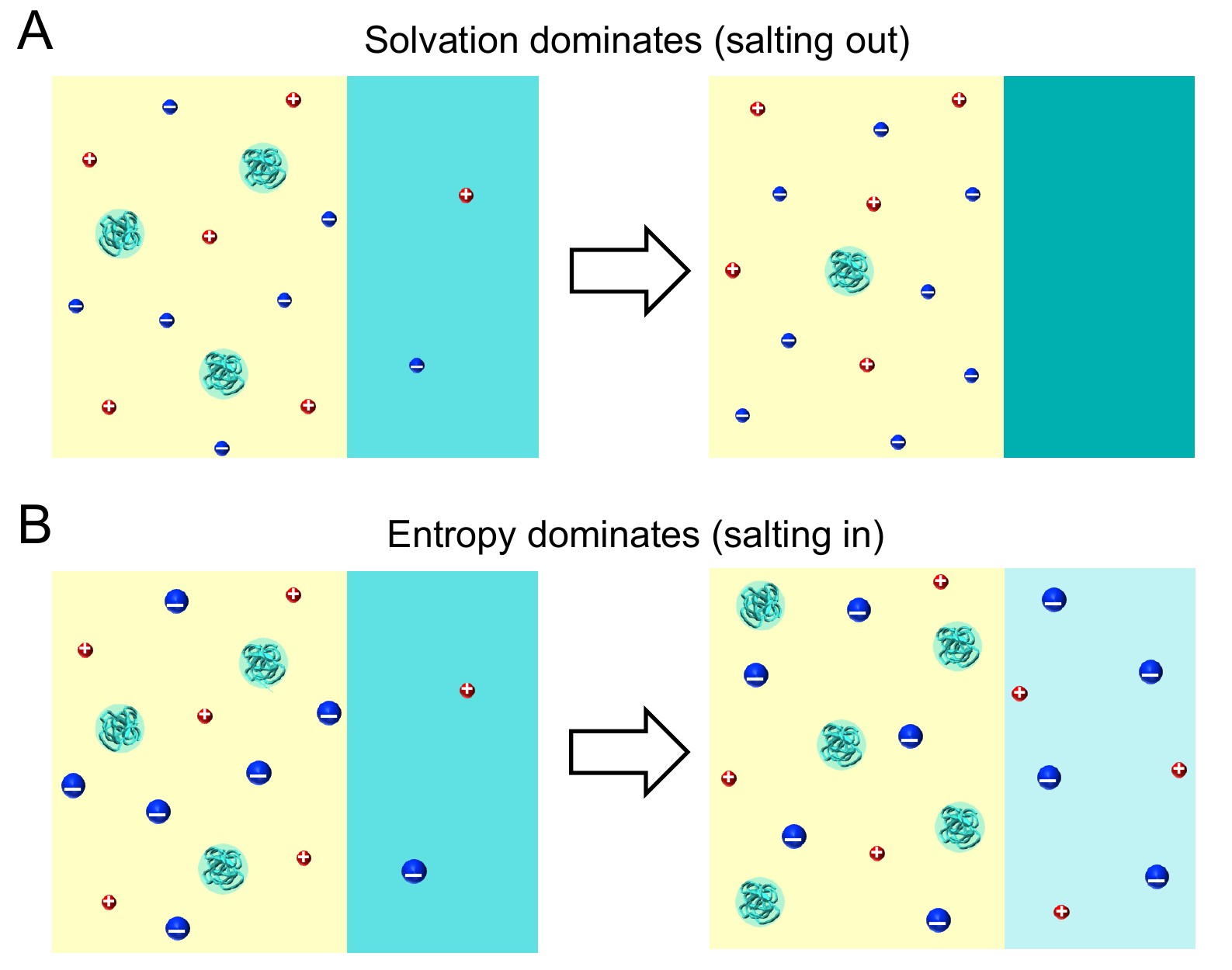}
\caption{Schematics of salt effects on the LLPS of proteins in the high salt concentration regime. ({\it A}) Ion solvation dominates for the case of small ions, which favors salting-out. ({\it B}) Translational entropy of ion dominates for the case of large ions, which favors salting-in.}
\label{fig4}
\end{figure}

Based on the electrostatic contributions to the chemical potential in Eq. 9, the competition between ion solvation and translational entropy can be quantified by:
\begin{align}
\Delta{\mu}^{elec}_P
= & {\mu}^{elec}_P (D) -{\mu}^{elec}_P (C) \nonumber\\
\approx & (z_+ + z_-)Nv_0c_b \left[\frac{l_{B,0}}{2} \left(\frac{\epsilon_S-\epsilon_P}{\epsilon^2_S}\right)\frac{1}{\bar{a}} - 1 \right] \tag{10}
\end{align}
where $\bar{a}$ is the valency-weighted harmonic average radius of cation and anion given by $(z_+ + z_-)/\bar{a}=z^2_+z_-/a_+ + z^2_-z_+/a_-$. $l_{B,0}=e^2/(4\pi\epsilon_0kT)$ is the Bjerrum length in vacuum. The detailed derivation of Eq. 10 is provided in SI, Section 2. $\Delta{\mu}^{elec}_P$ represents the driving force for a single protein to transfer from the concentrated phase (Phase C) to the dilute phase (Phase D). $\Delta{\mu}^{elec}_P>0$, ion solvation dominates, and protein prefers to stay in the concentrated phase rather than the dilute phase, which indicates salting-out. $\Delta{\mu}^{elec}_P<0$, translational entropy dominates, indicating salting-in. Equation 10 shows that the solvation effect becomes less pronounced as $\bar{a}$ increases. This explains our numcerical results in Fig. 2 and the experimental observations that protein salting-in occurs for larger ions.
This can also explain the specific ion effect that protein solubility increases with the anion radius, consistent with the trend of direct Hofmeister series observed in the high salt concentration regime \cite{Zhang2009,Cho2008,Mason2010,Otis2022,Zhang2005,Weckstrom2007,Shimada2011}.
Furthermore, the solvation energy depends on ion valency as well. From the expression of $\bar{a}$, ions with higher valency can be equivalently interpreted as monovalent ions with smaller effective radius. Therefore, multivalent ions promote salting-out. It explains the experimental findings in various protein and polymer solutions that SO$^{2-}_4$ shows much stronger tendency of salting-out even than Cl$^{-}$, although $a_{\rm {SO^{2-}_4}}$ is larger than $a_{\rm {Cl^-}}$ \cite{Cho2008,Otis2022,Zhang2005}.

As indicated by Eq. 10, the solubility at high salt concentrations also depends on the dielectric constant of protein $\epsilon_P$.  $\Delta{\mu}^{elec}_P$ decreases with the increase of $\epsilon_P$, preferring salting-in. This is consistent with the experimental observation that lysozyme with higher $\epsilon_P$ has stronger tendency of salting-in than elastin-like polypeptide with lower $\epsilon_P$.
Baldwin measured the solubility of peptide and observed that salting-out becomes more pronounced as the number of hydrocarbon side groups increases \cite{Baldwin1996}. More hydrocarbon side groups leads to the reduction of the dielectric constant of peptide. Furthermore, Shimada et al. recently investigated the LLPS of ureido-derivatized polymers \cite{Shimada2011}. They found that the solubility behavior turns from salting-out to salting-in as more ureido groups are grafted the polymer. The ureido group is highly polar and hence expected to increase the dielectric constant of the polymer \cite{Wyman1933,Klotz1996}. Their experimental results can be well captured by our theory.

\begin{figure}[h]
\centering
\includegraphics[width=0.45\textwidth]{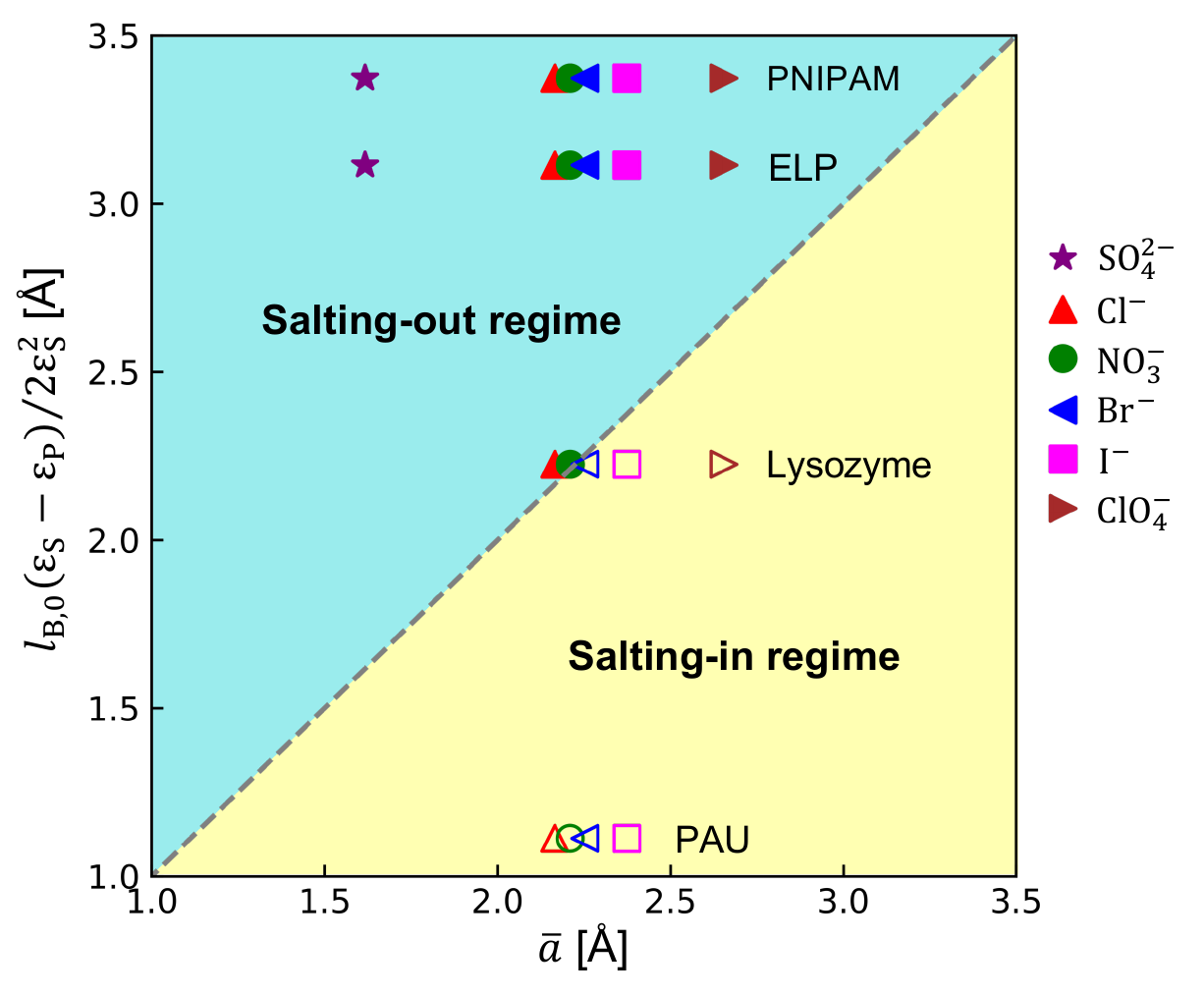}
\caption{Comparison of the solubility behavior predicted by our theory with experimental results for various proteins and polymers in concentrated sodium solutions with different anions. The dash diagonal line is the universal criterion given by Eq. 11 for determining the boundary between the salting-in and salting-out regimes. Scattering data points represent the experimental results reported in literature, where open and filled symbols denote salting-in and salting-out behaviors, respectively.
The Born radii of Cl$^-$, NO$^-_3$, Br$^-$, I$^-$, ClO$^-_4$, SO$^{2-}_4$ and Na$^+$ are 1.91, 1.98, 2.05, 2.26, 2.83, 3.79 and 2.5 \text{\normalfont\AA}, respectively \cite{Levin2010}.
The dielectric constant of water $\epsilon_S$=80. The dielectric constants of lysozyme \cite{Smith1993}, elastin-like polypeptide (ELP)\cite{Baer2006}, PNIPAM\cite{Rullyani2020} and poly(allylamine)-copoly(allylurea) (PAU)\cite{Ureido_Note} are 30, 10, 4.2, 56, respectively.
}
\label{fig5}
\end{figure}

Our theory provides a simple analytical criterion for determining the solubility behavior, i.e. salting-in versus salting-out. From $\Delta{\mu}^{elec}_P=0$ in Eq. 10, the boundary between the salting-in and salting-out regimes is given by the following universal line:
\begin{align}
\frac{l_{B,0}}{2} \left(\frac{\epsilon_S-\epsilon_P}{\epsilon^2_S}\right) =& \bar{a}  \tag{11}
\end{align}
This analytical result in Eq. 11 is confirmed by the numerical calculations (Fig. S1). To directly compare our theoretical predictions with experimental measurements, Fig. 5 shows the solubility behaviors of two proteins (lysozyme \cite{Zhang2009} and elastin-like polypeptide \cite{Cho2008}) and two synthetic polymers (PNIPAM \cite{Zhang2005} and poly(allylamine)-copoly(allylurea) \cite{Shimada2011}) in the solutions of sodium salts with various anions. For a specific pair of protein and anion, salting-in result observed in experiment is denoted by open symbol whereas salting-out result is denoted by filled symbol. These two types of data points locate almost exactly within the corresponding regimes separated by the universal line predicted by Eq. 11. 
The protein solubility increases following SO$^{2-}_4$, Cl$^-$, NO$^-_3$, Br$^-$, I$^-$ and ClO$^-_4$, precisely the direct Hofmeister  series \cite{Zhang2010,Okur2017,Nostro2012}. It is clear that our theoretical result is quite universal, which captures the known salt concentration effect and specific ion effect on LLPS of different proteins and polymers.

It should be noted that our prediction is in quantitative agreement with experimental results without any fitting parameters. Only intrinsic parameters of ions such as the valency and radius as well as the dielectric constant of protein are needed, which can either be adopted from literature or measured in experiments.  It is also interesting to note that our theory captures the salt effects on LLPS by only considering the contribution of Born energy in the ion solvation, indicating its dominant role for simple ions like halogen anions. However, for ions with more complex constitutions and structures, other contributions such as hydration, dispersion and polarization should also be taken into account \cite{Levin2010,Nostro2012,Wang2014,Salis2014}.
%\textcolor{red}{This explains the discrepancy between theory and experiment for the data of ClO$^-_4$ in elastin-like polypeptide as shown in Fig. 5.}
However, the existence and relative importance of these higher order effects on LLPS can only be evaluated when the essential Born solvation energy and translational entropy of ions are systematically treated as in our theory.

\subsection*{Protein Solubility vs. Charge Inversion} The preferable distribution of ions in the high dielectric medium also leads to counterintuitive surface properties of protein. Salis et al. observed that the electrophoritic mobility of a single lysozyme changes sign as salt concentration increases, indicating an inversion of the effective surface charge of the protein \cite{Salis2012}. Bostr$\ddot{\rm o}$m et al., suggested that charge inversion can explain the specific ion effect on protein solubility: reversal of the Hofmester series from the inverse sequence at low salt concentrations to the direct sequence at high salt concentrations \cite{Bostrom2011}. Similar charge inversion can also be predicted by our theory. Figure 6{\it A} shows the electrostatic potential profiles $\psi(r)$ of a single protein in the presence of various anions. As anion radius increases, $\psi(r)$ turns from its original positive value to negative, which indicates the occurrence of charge inversion for larger anions. This is in agreement with previous experimental results and theoretical predictions. The charge inversion can be explained by the induced Galvani potential and local charge separation at protein surface due to different Born solvation energy between cation and anion \cite{Wang2011}. The induced negative charge becomes stronger with the increase of $a_{-}$, which can even overcompensate the original positive charge of protein as the net charge density shown in the inset of Fig. 6{\it A}. It is worth noting that charge inversion is usually observed in the presence of multivalent counterions as a result of strong ionic correlation. Here, it occurs in monovalent salt systems, which is driven by the unequal solvation energy between cation and anion.

\begin{figure}[h]
\includegraphics[width=0.45\textwidth]{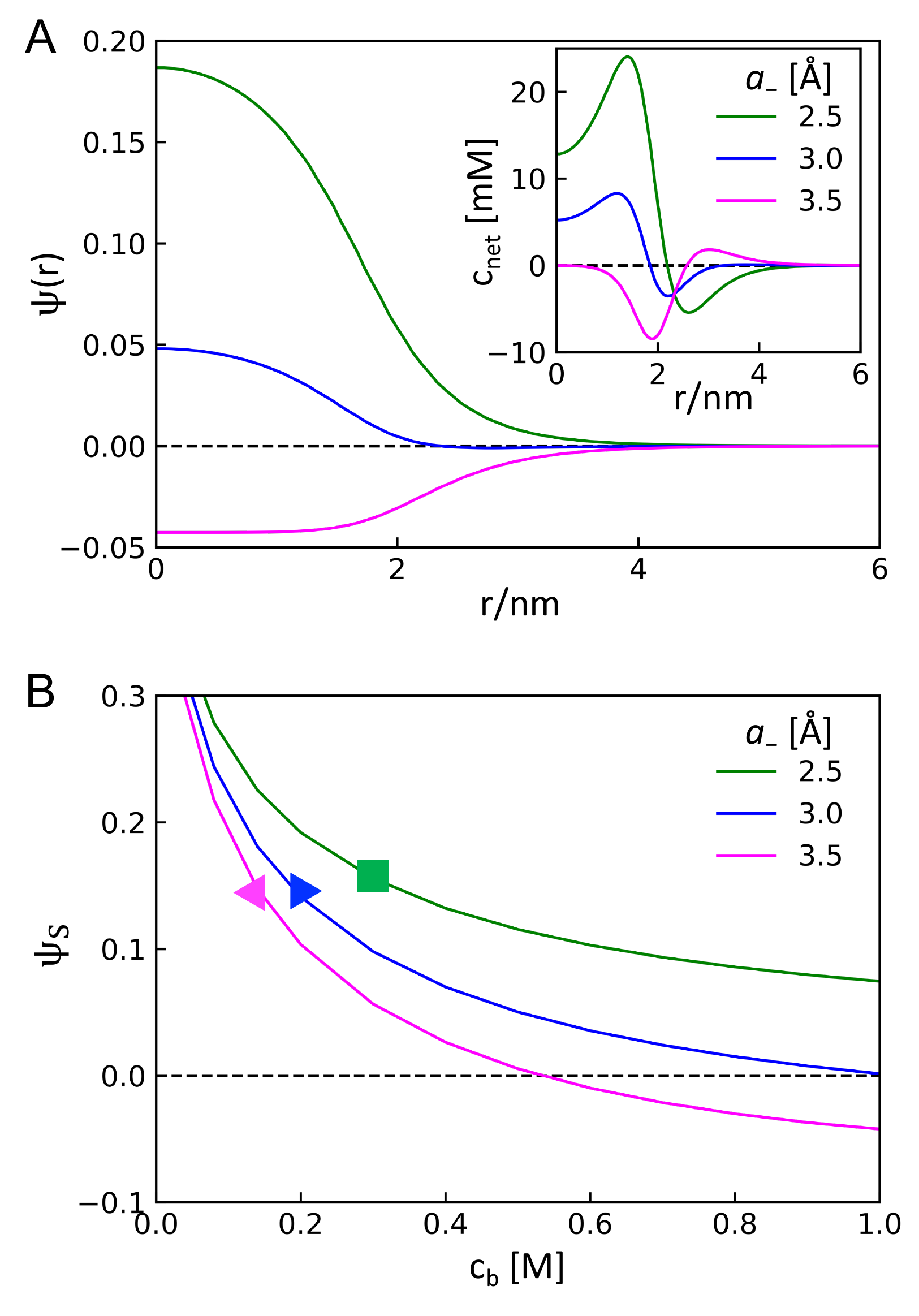}
\caption{The relation between charge inversion and solubility. ({\it A}) Electrostatic potentials $\psi(r)$ profile of a single protein in the presence of various anions.  $a_+=2.5\mathring{\rm A}$, $z_+=z_-=1$, $\epsilon_P=30$, and $c_b=0.7$M. The inset shows the profiles of the net charge density $c_{\rm net}(r)=z_+c_+(r)-z_-c_-(r)+\alpha{\rho}_P(r)/v_0$.
({\it B}) Electrostatic potentials at the protein surface $\psi_{S}$ as a function of salt concentration $c_b$. Filled symbols in Fig. 6{\it B} locate the corresponding turning points from salting-out to salting-in shown in Fig. 2{\it A}. }
\label{fig6}
\end{figure}

We further compare the charge inversion for a single protein and the collective solubility behavior for the entire protein solution to elucidate their relation. Figure 6{\it B} plots the electrostatic potential at protein surface $\psi_{S}$ as a function of salt concentration. Charge inversion can only been observed for larger anion with $a_-=3.5\text{\normalfont\AA}$. By comparing the results of Fig. 2{\it A} and Fig. 6{\it B}, it can been clearly seen that the turning point from salting-out to salting in for $a_-=3.5\text{\normalfont\AA}$ appears at $c_b=0.16$M, which is different from the charge inversion point $c_b=0.55$M. The inconsistency is more obvious for smaller ions ($a_-=2.5\text{\normalfont\AA}$ and $a_-=3.0\text{\normalfont\AA}$), where the turning points of solubility can be observed in Fig. 2{\it A} but the charge inversion is absent within the given range of salt concentration. Similar inconsistency can also been found from the comparison between the electrophoritic mobility of a single protein and the solubility of protein solutions measured by experiments \cite{Zhang2009,Salis2012}. For the same lysozyme protein, the charge inversion occurs for all anions, whereas on the contrary, the turning of the solubility from salting-out to salting-has not been observed for Cl$^{-}$. Furthermore, charge inversion is also inconsistent with the reversal of Hofmeister series. Figure 2{\it A} shows that the reversal of Hofmeister series from the inverse sequence to the direct sequence occurs around $c_b=0.25$M, largely different from the charge inversion point shown in Fig. 6{\it B}. Therefore, our theoretical results suggest that the counterintuitive behaviors of turning solubility and reversal of Hofmeister series observed at high salt concentrations are not related to charge inversion. They are actually attributed to the competition between the ion solvation and translational entropy as elucidated in the above subsection.

\section{Conclusions and Discussion}

We develop a self-consistent theory to study salt effect on LLPS of protein solutions by systematically incorporating electrostatic interaction, hydrophobicity, ion solvation and transnational entropy into a unified framework. Our theory has made important improvements compared to previous mean-field work. Both the highly localized density fluctuation of proteins in the dilute phase and the electrostatic fluctuation (manifested by the self-energy of ions) are explicitly accounted for. The long-standing puzzles of the non-monotonic salt concentration dependence and the specific ion effect are fully captured by our theory. We find that proteins show salting-out at low salt concentrations due to ionic screening. The solubility decreases with the increase of anion radius, following the inverse Hofmeister series. On the other hand, in the high salt concentration regime, protein remains salting-out for small ions but turns to salting-in for larger ions. The Hofmeister series is reversed to the direct sequence. We reveal that both the turning of solubility from salting-out to salting-in and the reversal of the Hofmeister series are attributed to the competition between the solvation energy and translational entropy of ions, but are not related to charge inversion of a single protein. Furthermore, we derive an analytical criterion for determining the boundary between the salting-in and salting-out regimes. Without any fitting parameters, the theoretical prediction is in quantitative agreement with experimental results for various proteins and polymers in sodium solutions with a broad range of anions.

Our theory reveals the essential physical chemistry of salt effects on LLPS using a simple charged macromolecular model, which can also be applied to other soft matter systems. The theory can be generalized to macromolecules with more complicated structures (e.g., chain architecture, heterogenous composition and charge distribution, local rigidity, helicity, etc.) and interactions that better represent real proteins. Although the charged macromolecular model seems only applicable to unfolded or intrinsic disordered proteins, the salt concentration effect and specific ion effect elucidated here is universal for both the unfolded and folded proteins. This is because the description of a giant liquid-like condensate is not sensitive to the folding details of a single protein. Furthermore, our theory captures the salt effects on LLPS by only considering the contribution of Born energy in the ion solvation, indicating its dominant role for simple ions like halogen anions. However, other contributions such as hydration, dispersion and polarization should also be taken into account for ions with more complex structures. These effects can be straightforwardly incorporated into the current theoretical framework. The existence and relative importance of these higher order effects on LLPS can only be evaluated when the essential Born energy and translational entropy of ions are accurately treated as in our work. The fundamental insight revealed here provides important guidance for modulating the LLPS of proteins via the addition of salt as an effective tool, which helps understand the functions of cellular organization and rationally design therapy for diseases.

\section{Materials and Methods}
We use spherical coordinate in the numerical calculations of self-consistent field theory based on the symmetry of spherical aggregate. Both the protein density and electrostatic potential are set to be zero at the boundary of the spherical simulation box.
The Crank–Nicolson method is used to solve the modified diffusion equation of chain propagator (Eq. (2)) \cite{Hoffman2001}.
The number of points that the chain contour has been discretized is set to be $N_s=1000$. The grid lattices are set such that the lattice spacings are smaller than 0.1$b$.
$\mu_S$ is set to be $-1$, such that the free energy of the reservoir of pure salt solution outside of the subvolume is 0. The equilibrium structure and the free energy can be obtained by solving Eqs. (1)-(3) iteratively until convergence. To accelerate the convergence, we use the following strategy to update the fields. Fields  conjugate to  the density of protein and solvent molecules are updated by a simple mixing rule, i. e., $\omega^{new}_{P,S} \leftarrow \lambda \omega^{new}_{P,S} +(1-\lambda) \omega^{old}_{P,S}$.
The same rule is adopted for updating electrostatic potential $\psi$ and Born energy $u_{\pm}$. 
The field conjugated to the incompressibility condition is updated by $\eta^{new} \leftarrow \eta^{old}+\kappa (\rho_P+\rho_S-1)$, where the second term on the r.h.s is adopted to reinforce the incompressibility.  $\lambda$=0.01 and $\kappa$=2.0 are chosen in our calculations. The relative errors for the free energy and the incompressibility condition are set to be below $10^{-11}$ and $10^{-7}$, respectively.

\section{Acknowledgment}
Acknowledgment is made to the donors of the American Chemical Society Petroleum Research Fund for partial support of this research. This research used the computational resources provided by the Kenneth S. Pitzer
Center for Theoretical Chemistry.

\end{document}